\documentclass[12pt]{article}

\advance\voffset by -2.0cm \advance\hoffset by -1.25cm
\usepackage{amsmath}
\usepackage{amssymb}
\usepackage{amsthm}
\usepackage{amsfonts}
\usepackage{url}
\usepackage{hyperref}

\theoremstyle{definition}

\newcommand{\CC}{\mathbb{C}} 

\newcommand{\be}{\begin{equation}}
\newcommand{\ee}{\end{equation}}

\newlength{\oldcolsep}

\numberwithin{equation}{section}

\begin{document}

\title{Universe as Klein-Gordon Eigenstates}
\author{Marco Matone}\date{}

\maketitle

\begin{center} {\it Dipartimento di Fisica e Astronomia ``G. Galilei''} \\
{\it Universit\`a di Padova, Via Marzolo 8, 35131 Padova,
Italy}

\vspace{.6cm}

{\it INFN, Sezione di Padova} \\
{\it Via Marzolo 8, 35131 Padova, Italy} \\
\end{center}

\begin{abstract}
\noindent
We formulate Friedmann's equations as second-order linear differential equations. This is done using
techniques related to the Schwarzian derivative that selects the $\beta$-times $t_\beta:=\int^t a^{-2\beta}$, where $a$
is the scale factor.
In particular, it turns out that Friedmann's equations are equivalent to the eigenvalue problems
$$
O_{1/2} \Psi=\frac{\Lambda}{12}\Psi \ , \qquad O_1 a =\frac{\Lambda}{3} a \ ,
$$
which is suggestive of a measurement problem.
$O_{\beta}(\rho,p)$ are space-independent Klein-Gordon operators,
depending only on energy density and pressure, and related to the Klein-Gordon
Hamilton-Jacobi equations. The $O_\beta$'s are also independent of the spatial curvature, labeled by $k$, and absorbed in
$$
\Psi=\sqrt a e^{\frac{i}{2}\sqrt{k}\eta} \ .
$$
The above pair of equations is the unique possible linear form of Friedmann's equations unless $k=0$, in which case
there are infinitely many pairs of linear equations. Such a uniqueness
just selects the conformal time $\eta\equiv t_{1/2}$ among the $t_\beta$'s, which is the key to absorb the curvature term.
An immediate consequence of the linear form is that it reveals a new symmetry of Friedmann's equations in flat space.

\end{abstract}

\newpage

\section{Introduction}

Unlike other fundamental laws of Nature, Einstein's field equations \cite{Einstein:1916vd}
\begin{equation}
R_{\mu\nu}-\frac{1}{2}Rg_{\mu\nu}+\Lambda g_{\mu\nu}=\frac{8\pi G}{c^4}T_{\mu\nu} \ ,
\label{Einstein}\end{equation}
are non-linear.
A great simplification of Eq.(\ref{Einstein}) has been derived by Friedmann \cite{Friedmann:1922,Friedmann:1924bb}, whose equations
are fundamental in modern cosmology. These equations encode the essence of general relativity,
therefore they are crucial not only for understanding the dynamics
of the Universe, but also to deepen other fundamental questions of Nature, such as
relations, yet to be discovered, between general relativity
and quantum mechanics, which is described by a linear equation.

\vspace{.2cm}

\noindent
A problem with Friedmann's equations is that, despite the simplification of Eq.(\ref{Einstein}),
the first one is highly non-linear.
In particular,
due to the term $k/a^2$, the problem seems particularly stringent in the
case of non flat space.
On the other hand, non-linearity is the main obstacle for a quantization of gravity.
Non-linearity of the first Friedmann equation is also one of the reasons why a more systematic analysis
of Friedmann's equations in their general form is still lacking, and they
are frequently considered in the simplest case of barotropic fluids, with a constant barotropic index.
It is then clear
that a linear form of Friedmann's equations would be of considerable interest in several contexts.

\vspace{.2cm}

\noindent
Here we show that the structure of Friedmann's equations hides an underlying linearity and can be formulated in the form
of a pair of linear second-order differential equations. It turns out that the two linear equations
correspond to the unique possible linear form, unless $k=0$,
that gives infinitely many pairs of linear versions of Friedmann's equations. Remarkably, as we will see,
an immediate consequence of the linear form is that it reveals a new symmetry of Friedmann's equations in flat space.
Another feature of the formulation is that the uniqueness of the linear form in the case $k\neq 0$ naturally selects
the conformal time and leads to a reformulation of Friedmann's equations in the form of two coupled space-independent Klein-Gordon equations.
Such operators are also independent of space curvature that, in this formulation, is absorbed in the Klein-Gordon wave-function.
The linear form of Friedmann's equations also shows that the cosmological constant plays the role of eigenvalues of the Klein-Gordon
operators. As a result, the outcome resembles a measurement problem so that suggesting a possible relation with multiuniverse theory.

\vspace{.2cm}

\noindent
The above analysis is related to the identification of the cosmological constant with the Wheeler-DeWitt quantum potential,
that plays the role of intrinsic energy \cite{Faraggi:2020blm}, and that as been derived in the context of the geometrical formulation of the
quantum Hamilton-Jacobi theory \cite{Faraggi:1996rn}-\cite{Faraggi:2012fv}. Even in the present investigation the two Klein-Gordon equations
are related to the quantum Hamilton-Jacobi theory and, in particular, there is a relation between their associated quantum potentials.

\vspace{.2cm}

\noindent
The strategy of the present investigation is to first introduce linear combinations of Friedmann's equations, with the condition
that they can be expressed in terms
of Schwarzian derivatives. This naturally introduces the $\beta$-times $t_\beta:=\int^t a^{-2\beta}$.
Then, using the chain rule for the Schwarzian derivative, the curvature term $k/a^2$ is absorbed by exponentiating the conformal time $\eta\equiv t_{1/2}$.
The result is a Schwarzian equation
for $e^{i\sqrt{k}\eta}$. On the other hand, any Schwarzian's equation is solved by the ratio of two linearly independent solutions
of the associated second-order linear differential equation. Together with the second Friedmann equation, such an equation provides
the promised linear form.

\section{Schwarzian form of Friedmann's equations}

Let us consider the Friedmann-Lema\^{\i}tre-Robertson-Walker (FLRW) line element
\begin{equation}
ds^2=-dt^2+a^2(t)\Big[\frac{dr^2}{1-kr^2}+r^2(d\theta^2+\sin^2\theta d\phi^2)\Big] \ ,
\end{equation}
where $k$ is either $+1$, $0$ and $-1$ for spatially closed, flat and open universes,
respectively. In the FLRW background the curvatures read
\begin{align}
R_0^{\;\;0} &=\frac{3\ddot a}{a} \ , \cr
R_j^{\;\;k} &=\Big(\frac{\ddot a}{a}+\frac{2\dot a^2}{a^2}+\frac{2k}{a^2}\Big) \delta_j^{\;\;k} \ , \cr
R & =6\Big(\frac{\ddot a}{a}+ \frac{\dot a^2}{a^2}+\frac{k}{a^2}\Big) \ ,
\label{riportaR}\end{align}
and the spatial scalar curvature is
\begin{equation}
^3R=6\frac{k}{a^2} \ .
\end{equation}
Assuming that the energy momentum tensor is the one of an ideal perfect fluid
\begin{equation}
T_\mu^{\;\;\nu}=\text{Diag}(-\rho,p,p,p) \ ,
\end{equation}
with $\rho$ and $p$ the energy density and pressure density, respectively, we get, from Einstein's equations, Friedmann's equations
\begin{equation}
\frac{{\dot a}^2}{a^2} = \frac{1}{3}(8\pi G\rho+\Lambda)-\frac{k}{a^2} \ ,
\label{F1}\end{equation}
and
\begin{equation}
\frac{\ddot a}{a} = -\frac{4\pi G}{3}(\rho+3p)+\frac{\Lambda}{3} \ ,
\label{F2}\end{equation}
where $\Lambda$ is the cosmological constant. Taking the time derivative of (\ref{F1}) and replacing $\ddot a/a$ by the right-hand side of (\ref{F2}) leads to the
continuity equation
\begin{equation}
\dot\rho+3H(\rho+p)=0 \ ,
\label{continuity}\end{equation}
with $H=\dot a/a$ the Hubble parameter.

\noindent
Friedmann's equations have two nice properties that follow by considering the linear combination of their left-hand sides
\begin{equation}
X_\beta(a):=\frac{\ddot a}{a}+(\beta-1)\left(\frac{\dot a}{a}\right)^2 \ .
\label{betauno}\end{equation}
The first property is that $X_\beta$ satisfies the relation
\begin{equation}
(\beta-\gamma)X_\alpha(a)+(\gamma-\alpha)X_\beta(a)+(\alpha-\beta)X_\gamma(a)=0 \ .
\label{Xrelation}\end{equation}
Furthermore, $X_\beta$ vanishes for $a=t^{1/\beta}$.
Both properties are related to the M\"obius symmetry of the Schwarzian derivative.
To show this let us first introduce the $\beta$-times, $t_\beta$, by
\begin{equation}
\dot t_\beta:=a^{1/\delta(\beta)} \ ,
\end{equation}
with $\delta(\beta)$ fixed by requiring that
\begin{equation}
X_\beta(a)=\delta\left[\frac{\dddot t_\beta}{\dot t_\beta}+(\delta\beta-1)\left(\frac{\ddot t_\beta}{\dot t_\beta}\right)^2 \right] \ ,
\label{betadue}\end{equation}
be proportional to the Schwarzian derivative of $t_\beta$
\begin{equation}
\{t_\beta,t\}=\frac{\dddot t_\beta}{\dot t_\beta}-\frac{3}{2}\left(\frac{\ddot t_\beta}{\dot t_\beta}\right)^2 \ .
\label{schwarzofz}\end{equation}
This condition gives $\delta=-1/(2\beta)$ so that
\begin{equation}
t_\beta(t)=\int_{0}^{t}dt' a^{-2\beta} \ .
\label{zetaea}\end{equation}
We then have $X_\beta(a)=-\left\{ t_\beta,t\right\}/(2\beta)$, that is
\begin{equation}
\frac{\ddot a}{a}+(\beta-1)\left(\frac{\dot a}{a}\right)^2=-\frac{1}{2\beta}\{t_\beta,t\} \ .
\label{28}\end{equation}
Since the Schwarzian derivative (\ref{schwarzofz}) is invariant under the ${\rm PSL}(2,\CC)$ linear fractional transformations
\begin{equation}
t_\beta\, \longrightarrow \, t_\beta'=\frac{A t_\beta+B}{Ct_\beta+D} \ ,
\label{induces}\end{equation}
it follows that
\begin{equation}
X_\beta(a')=X_\beta(a) \ ,
\end{equation}
where $a'$ is the transformation of $a$ induced by (\ref{induces}), that is (we set $AD-BC=1$)
\begin{equation}
a \, \longrightarrow \,  a'=a({Ct_\beta+D})^{1/\beta} \ .
\label{cometrasformaa}\end{equation}
Let us consider the identity
\begin{equation}
\dot t_\beta^{1/2}\frac{d}{d t}\frac{1}{\dot t_\beta}\frac{d}{d t}\dot t_\beta^{1/2}\psi_\beta=\Bigg(\frac{d^2}{d t^2}
+\frac{1}{2}\{t_\beta,t\}\Bigg)\psi_\beta \ .
\label{impliestheid}\end{equation}
Inspection of the left-hand side shows that the linear span of the two linearly-independent functions
$\psi_\beta=\dot t_\beta^{-1/2}$ and $\psi_\beta^D=\dot t_\beta^{-1/2} t_\beta$
is the kernel of the operator in round brackets.
It follows that solving the Schwarzian equation $\{t_\beta,t\}=2U_\beta$ is equivalent to solve the associated second-order linear
differential equation
\begin{equation}
\Big(\frac{d^2}{d t^2}+U_\beta\Big)\phi_\beta=0 \ .
\label{KG}\end{equation}
In particular, since
\begin{equation}
\psi_\beta=a^{\beta} \ , \qquad \psi_\beta^D=a^{\beta}t_\beta \ ,
\label{solutions}\end{equation}
it follows that the solution of the Schwarzian equation is
\begin{equation}
t_\beta=\frac{\psi_\beta^D}{\psi_\beta} \ .
\label{ratio}\end{equation}
By construction it is clear that (\ref{ratio}) is still valid even if $\psi_\beta$ and $\psi_\beta^D$ are replaced
by their arbitrary linearly independent combinations
\begin{equation}
\begin{pmatrix}
{\psi_\beta^D}' \\
\psi_\beta'
\end{pmatrix}
=
\begin{pmatrix}
A & B \\
C & D
\end{pmatrix}
\begin{pmatrix}
\psi_\beta^D \\
\psi_\beta \
\end{pmatrix} \ ,
\label{LocalitaLePanche}\end{equation}
showing that such an arbitrariness corresponds to the M\"obius invariance of the Schwarzian derivative (\ref{induces}). This means that M\"obius transformations
are related to the boundary conditions of the Schwarzian and linear equations.
Let us stress that (\ref{impliestheid}) implies that expressing the time derivative in terms of $t_\beta$ trivializes the equation.
Actually, we have
\begin{equation}
\dot t_\beta^{1/2}\frac{d}{d t}\frac{1}{\dot t_\beta}\frac{d}{d t}\dot t_\beta^{1/2}\psi_\beta=\dot t_\beta^{3/2}\frac{d^2}{d t_\beta^2}\phi_\beta =0 \ ,
\label{lhsss}\end{equation}
so that $\phi_\beta=At_\beta+B$, where $A\neq0$ and $B$ are constants. This is a key point
for the trivialization of the quantum Hamilton-Jacobi equation \cite{Faraggi:2020blm}-\cite{Faraggi:2012fv}.

\noindent
The above analysis shows that Friedmann's equations can be expressed in terms of Schwarzian derivatives.
In particular, by (\ref{F1}), (\ref{F2}) and (\ref{28}), we have
\begin{equation}
\frac{1}{2}\{ t_\beta,t\}= V_\beta(\rho,p)-\beta^2\frac{\Lambda}{3} +\beta(\beta-1)\frac{k}{a^2} \ ,
\label{fried12}\end{equation}
where
\begin{equation}
V_{\beta}(\rho,p)=-\frac{4}{3}\pi G\beta\left[(2\beta-3)\rho-3p\right] \ ,
\label{Vbeta}\end{equation}
that satisfies the relation
\begin{equation}
\frac{V_\alpha}{\alpha}=\frac{\alpha-\gamma}{\beta-\gamma}\frac{V_\beta}{\beta}+\frac{\alpha-\beta}{\gamma-\beta}\frac{V_\gamma}{\gamma} \ .
\label{Vrelation}\end{equation}

\section{A new symmetry from the linear form}

Here we show that in the case of flat space, the linear form of Friedmann's equations reveals an interesting new symmetry.
Let us start by noticing that since for any pair $(\alpha,\beta)\in \CC^{2}\backslash\{0,0\}$, $\alpha\neq\beta$, $X_{\alpha}(a)$ is not proportional to  $X_{\beta}(a)$, it follows that
Friedmann's equations in flat space have the {\it canonical eigenvalue form}
\begin{equation}
\begin{pmatrix}
\displaystyle  O_\alpha & 0 \\
0 & \displaystyle O_\beta
\end{pmatrix} \Psi_{\alpha \beta}
=\frac{\Lambda}{3}\begin{pmatrix}
\displaystyle \alpha^2  & 0   \\
0  & \displaystyle \beta^2
\end{pmatrix}
\Psi_{\alpha \beta} \ ,
\label{generale}\end{equation}
where
\begin{equation}
\Psi_{\alpha \beta}=\begin{pmatrix}
a^\alpha t_\alpha \\
a^\alpha \\
a^\beta t_\beta\\
a^\beta
\end{pmatrix}  \ ,
\end{equation}
and
\begin{equation}
O_\beta(\rho,p):=\frac{d^2}{d t^2}+V_{\beta}(\rho,p) \ ,
\end{equation}
are the space-independent Klein-Gordon operators.
Also note that for each $\beta$ there is a dual canonical equation, the one with label $-\beta$, with has the same eigenvalue $\Lambda \beta^2/3$.
It follows that the set of all possible canonical forms (\ref{generale}) can be grouped in the eigenvalue problems
\begin{equation}
\begin{pmatrix}
\displaystyle  O_{-\beta} & 0 \\
0 & \displaystyle O_\beta
\end{pmatrix} \Psi_{-\beta \beta}
=\beta^2\frac{\Lambda}{3}
\Psi_{-\beta\beta} \ .
\label{eigenvalues}\end{equation}
The above shows that, in the case of flat space, there are infinitely main pairs
of linear differential equations which are equivalent to Friedmann's equations (\ref{F1}) and (\ref{F2}).
On the other
and, they are related in a non-linear way since are satisfied by different powers of $a$.
It is precisely this non-linearity that
implies a non-trivial symmetry. In fact, we can map one equation to another simply
taking powers of $a$. In particular, note that the map $a\to a^\alpha$ corresponds to
$t_\beta\to t_{\alpha\beta}$. By $\dot t_\beta=a^{-2\beta}$, we have $\dot t_{\alpha\beta}= (\dot t_\beta)^\alpha$, implying
\begin{equation}
\{t_{\alpha\beta},t\}=\alpha\{t_\beta,t\}-\frac{1}{2}\alpha(\alpha-1)\left(\frac{\ddot t_\beta}{\dot t_\beta}\right)^2 \ .
\end{equation}
This can be also checked by observing that such a relation is equivalent to the identity
\begin{equation}
V_{\alpha\beta}-(\alpha\beta)^2\frac{\Lambda}{3}=\alpha V_\beta-\alpha\beta^2\frac{\Lambda}{3}-\frac{1}{3}\alpha(\alpha-1)\beta^2(8\pi G\rho+\Lambda) \ .
\end{equation}
On the other hand, by
\begin{equation}
V_{\alpha\beta}(\rho,p)= V_\beta(\alpha^2 \rho, \alpha p-\alpha(\alpha-1)\rho) \ ,
\end{equation}
it follows that all the canonical eigenvalues equations (\ref{generale}) are related by
\begin{equation}
\begin{aligned}
& a & \longrightarrow \quad &  a^\alpha \ , \\
& \rho &\longrightarrow \quad & \alpha^2 \rho \ , \\
& p & \longrightarrow \quad & \alpha p - \alpha(\alpha-1)\rho \ , \\
& \Lambda & \longrightarrow \quad & \alpha^2\Lambda \ ,
\end{aligned}
\label{Jimi}\end{equation}
which are equivalent to the following transformations
\begin{equation}
\begin{aligned}
& a & \longrightarrow \quad & a^\alpha \ , \\
& \rho & \longrightarrow \quad & \alpha^2 \rho+(\alpha^2-1)\frac{\Lambda}{8\pi G} \ , \\
& p & \longrightarrow \quad &  \alpha p - \alpha(\alpha-1)\rho +(1-\alpha^2)\frac{\Lambda}{8\pi G} \ , \\
& \Lambda & \longrightarrow \quad & \Lambda \ .
\end{aligned}
\label{Frank}\end{equation}
Finally note that since any pair (\ref{generale}) is equivalent to Friedmann's equations, it follows that
(\ref{Jimi}) and (\ref{Frank}) are symmetries of the Friedmann equations with $k=0$.

\section{Absorbing the curvature by exponentiation}

Note that any pair of equations (\ref{fried12}) with distinct values of $\beta\in\CC\backslash\{0\}$ is equivalent to Friedmann's equations.
Furthermore, the chain rule of the Schwarzian derivative $\{f,x\}=(\partial_xy)^2\{f,y\}+\{y,x\}$ implies
\begin{equation}
\{t_\alpha,t_\beta\}=\left(\frac{d t_\gamma}{d t_\beta}\right)^{2}\{t_\alpha,t_\gamma\}+\{t_\gamma,t_\beta\} \ ,
\label{124}\end{equation}
which is equivalent to
\begin{equation}
X_{\alpha\beta} X_{\beta\gamma}+X_{\alpha\gamma}X_{\gamma\beta}=X_{\gamma\beta}X_{\beta\gamma} \ ,
\end{equation}
where $X_{\alpha\beta}:=\{t_\alpha,t_\beta\}$.
In the theorem of section 9.2 of \cite{Faraggi:1998pd} it has been shown that
a related cocycle condition determines the Schwarzian derivative. It is worth recalling that (\ref{124}) characterizes the stress tensor
of 2D CFT and the quantum potential \cite{Faraggi:1998pd}.

\noindent
An immediate consequence of Eq.(\ref{fried12}) is that the conformal time
\begin{equation}
\eta \equiv t_{1/2}=\int_{0}^{t}\frac{dt'}{a} \ ,
\end{equation}
satisfies the equation
\begin{equation}
\{\eta,t\} = \frac{4}{3}\pi G(2\rho + 3p) - \frac{\Lambda}{6} - \frac{k}{2}\dot \eta^2 \ .
\label{conformaltimeeq}\end{equation}
Now note that, for $k\neq0$, the chain rule of the Schwarzian derivative implies the identity
\begin{equation}
\{e^{i\sqrt k \eta},t\}=\{\eta,t\}+\frac{k}{2}\dot\eta^2 \ ,
\label{rte}\end{equation}
which is the relation used in the exponential map $z\to e^z$ in the radial quantization of 2D CFT.
It follows that the spatial curvature term in (\ref{conformaltimeeq}) can be absorbed by the exponentiation $\eta\to e^{\pm i\sqrt{k} \eta}$ in the Schwarzian derivative,
equivalent to
\begin{equation}
a \longrightarrow \frac{a}{\sqrt k} \exp(\mp i\sqrt k \eta) \ .
\end{equation}
We then have that for $k\neq0$
\begin{equation}
\{e^{\pm i\sqrt{k} \eta},t\} = \frac{4}{3}\pi G(2\rho + 3p) - \frac{\Lambda}{6} \ ,
\label{curvaturebyexp}\end{equation}
which is invariant under the M\"obius transformations
\begin{equation}
e^{i\sqrt{k} \eta} \longrightarrow \frac{Ae^{i\sqrt{k} \eta}+B}{Ce^{i\sqrt{k} \eta}+D} \ .
\label{solvethis}\end{equation}
It follows that one can solve the non-linear problem (\ref{conformaltimeeq}) by taking the logarithm of the solution of (\ref{curvaturebyexp}), which
in turn determines $a$. On the other hand,
equation (\ref{curvaturebyexp}) is equivalent to the eigenvalue problem
\begin{equation}
\left[\frac{d^2}{d t^2}+\frac{2}{3}\pi G(2\rho + 3p)\right]\psi=\frac{\Lambda}{12}\psi\ .
\label{lineareee}\end{equation}
A key point is that in the case $k\neq0$, this equation and (\ref{F2}) is the unique pair of equations that are
solution to the problem of finding a linear form of Friedmann's equations. The reason is that any other linear form
should correspond to
adding Eq.(\ref{F2}) to (\ref{lineareee}). On the other hand, this would break its linear form. Also note that a key point
is to absorb the curvature term by exponentiation. On the other hand, by the chain rule of the Schwarzian derivative, it follows that
the exponentiation of $t_\beta$ gives a term proportional to $\dot{t_\beta}^2$, so that
the curvature term can be obtained only by exponentiating $\eta$.
It is then interesting that the solution of the problem just selects
the conformal time among all $t_\beta$-times.

\noindent
Two linearly independent solutions of (\ref{lineareee}) are
\begin{equation}
\phi=\sqrt a\exp\left(-\frac{i}{2}\sqrt{k}\eta\right) \ , \qquad \phi^D=\sqrt a\exp\left(\frac{i}{2}\sqrt{k}\eta \right) \ ,
\label{lesoluzioni}\end{equation}
which are hyperbolic functions when $k=-1$. As a check note that replacing (\ref{lesoluzioni}) in (\ref{lineareee}) yields
\begin{equation}
\frac{1}{2}\left(\frac{\dot a}{a}\right)^2-\frac{\ddot a}{a}= \frac{4}{3}\pi G(2\rho + 3p) - \frac{\Lambda}{6} - \frac{k}{2a^2} \ .
\end{equation}
Linear combinations reproducing the solutions $\psi_{1/2}=\sqrt a$ and $\psi_{1/2}^D=\sqrt{a}\eta$
in the $k\to 0$ limit are
\begin{equation}
\psi=\sqrt a \cos\left(\frac{\sqrt k}{2}\eta\right) \ , \qquad \psi^D=2\frac{\sqrt a}{\sqrt k}\sin\left(\frac{\sqrt k}{2}\eta\right) \ ,
\end{equation}
so that
\begin{equation}
a=\psi^2+\frac{k}{4}{\psi^D}^2 \ .
\label{eccoqua}\end{equation}
Solving the eigenvalue problem (\ref{lineareee}) gives $\psi$ and $\psi^D$ as functions of $\rho$, $p$ and $\Lambda$, that
by (\ref{eccoqua}) fix $a$.

\noindent
Let us introduce the wave-function
\begin{equation}
\Psi= \sqrt a e^{\frac{i}{2}\sqrt{k}\eta} \ ,
\label{thewavefunction}\end{equation}
so that, for $k\neq -1$,
\begin{equation}
a=|\Psi|^2 \ .
\end{equation}
We then have that the Friedmann equations are equivalent to
the following coupled space-independent Klein-Gordon eigenvalue problems
\begin{equation}
O_{1/2}\Psi=\frac{\Lambda}{12}\Psi \ ,
\label{unmezzo}\end{equation}
\begin{equation}
O_{1}a=\frac{\Lambda}{3}a \ ,
\label{unoooo}\end{equation}
that resembles a measurement problem, so that suggesting a possible role of the multiuniverse theory.
Note that, depending on the boundary conditions, the solutions of (\ref{unmezzo})  will be a linear combination of $\Psi$ and of a linear independent one,
that is $\Psi^D=\sqrt a e^{-\frac{i}{2}\sqrt{k}\eta}$. In the case of (\ref{unoooo}) a solution which is linearly independent of $a$ is $a\int^t a^{-2}$.

\noindent
Note that for $k=0,1$, Eq.(\ref{unoooo}) can be equivalently written in the form
$O_{1}|\Psi|^2=\frac{\Lambda}{3}|\Psi|^2$.
Let us write (\ref{unmezzo}) and (\ref{unoooo}) in the explicit form
\begin{equation}
\left[\frac{d^2}{d t^2}+\frac{2}{3}\pi G(2\rho + 3p)\right]\Psi=\frac{\Lambda}{12}\Psi \ ,
\label{lineareeeBISSS222}\end{equation}
\begin{equation}
 \left[\frac{d^2}{d t^2} +\frac{4\pi G}{3}(\rho+3p)\right]a=\frac{\Lambda}{3}a \ .
\label{F2BISSS222}\end{equation}
Eq.(\ref{eccoqua}) shows that the $k$-dependence of $a$ is determined by the initial conditions for
$\Psi$.
In this respect, observe that while the left-hand side of (\ref{rte}) is invariant under M\"obius transformations of $e^{i\sqrt k \eta}$, both terms in the right-hand side
undergo non-trivial transformations, an issue related to the quantum Hamilton-Jacobi equation. To show this, let us first define
\begin{equation}
Q[f]:= \frac{1}{f}\frac{d^2 f}{d t^2}=-\frac{1}{2}\left\{\int^t f^{-2},t\right\}  \ .
\label{wow0bis}\end{equation}
Note that (\ref{conformaltimeeq}), namely
\begin{equation}
-\frac{k}{4}{\dot\eta}^2 +\frac{2}{3}\pi G(2\rho + 3p) - \frac{\Lambda}{12}
-\frac{1}{2}\{\eta,t\} = 0 \ ,
\end{equation}
can be interpreted as the space-independent Klein-Gordon Hamilton-Jacobi equation \cite{Bertoldi:1999zv} associated to (\ref{lineareeeBISSS222}), with
\begin{equation}
Q[|\Psi|]=
 - V_{1/2}(\rho,p)+\frac{\Lambda}{12}+\frac{\delta_{k1}}{4 |\Psi|^4}\ ,
\label{wow0}\end{equation}
playing the role of quantum potential. Since the curvature term does not
contribute to (\ref{wow0}) when $\Psi$ in (\ref{thewavefunction})
is real, it follows that, for $k=-1$ and $k=0$, we can replace $\Psi$ in (\ref{lineareeeBISSS222}) by $|\Psi|$, so that
\eqref{wow0} coincides with (\ref{lineareeeBISSS222}).
In this respect note that, depending on the boundary conditions, one can consider real solutions of (\ref{lineareeeBISSS222}) by taking linear combinations
of $\Psi$ and $\Psi^D$. This is related to the Einstein paradox \cite{Faraggi:1998pd}.
We note that since $a$ is real, it follows that also its quantum potential has the same content of (\ref{F2BISSS222})
\begin{equation}
Q[a]=-V_1(\rho,p)+\frac{\Lambda}{3} \ .
\label{wow1}\end{equation}
Finally, we show that (\ref{F2BISSS222}) fixes the quantum potential associated to (\ref{lineareeeBISSS222}).
The fastest way to proceed is to first consider the case $k\neq -1$, so that $a=|\Psi|^2$ and we can then use the identity
\begin{equation}
Q[|\Psi|]=\frac{1}{2}\left[Q[|\Psi|^2] - \frac{1}{2|\Psi|^4}\left(\frac{d|\Psi|^2}{d t} \right)^2\right] \ ,
\end{equation}
that is
\begin{equation}
- V_{1/2}(\rho,p)+\frac{\Lambda}{12}+\frac{k}{4 a^2}=\frac{1}{2}\left[-V_1(\rho,p)+\frac{\Lambda}{3}
- \frac{1}{2}\frac{\dot a^2}{a^2}\right] \ ,
\end{equation}
then note that it extends to the case $k=-1$ because coincides
with Eq.(\ref{F1}).

\section*{Acknowledgements} It is a pleasure to thank  D. Bertacca,  K. Lechner and A. Ricciardone for interesting comments and
discussions.

\end{document}